\documentclass[APS,superscriptaddress,twocolumn]{revtex4-2}
\usepackage[colorlinks]{hyperref}
\usepackage{amsmath,amssymb,graphicx,color,amsfonts,hyperref,cancel}
\usepackage{comment}

\begin{document}
\title{Parametric instability of ultracold Bose gases with long-range interaction and quantum fluctuations trapped in optical lattices}

\author{Etienne Wamba}
\email {ettienne.wamba@ubuea.cm}
\affiliation{Faculty of Engineering and Technology, University of Buea, P.O. Box 63 Buea, Cameroon}
\affiliation{STIAS, Wallenberg Research Centre at Stellenbosch University, Stellenbosch 7600, South Africa}

\author{J\'er\'emie Bai}
\email {baijeremie9@gmail.com}
\affiliation{Department of Physics, Faculty of Science, University of Maroua, P. O. Box: 814 Maroua, Cameroon}

\author{ Aur\'{e}lien S. Tchakoutio Nguetcho}
\email {nguetchoserge@yahoo.fr}
\affiliation{Department of Physics, Faculty of Science, University of Maroua, P. O. Box: 814 Maroua, Cameroon}

\affiliation{ImViA Laboratory EA 7535, BP 47870, 21078 Dijon Cedex, Université de Bourgogne, France}

\affiliation{University of Duisburg-Essen, Chair for Nonlinear Analysis and Modelling, Faculty of Mathematics, Thea-Leymann-Stra{\ss}e 9, D-45127 Essen, Germany}

\date{\today}%

\begin{abstract}

%\texttt{The Abstract goes here .....} 

We investigate the dynamical instabilities of an ultracold Bose-Bose mixture with long-range dipole-dipole interactions, trapped in deep optical lattices and subject to periodically varying contact interaction. The effect of beyond-mean-field corrections due to quantum fluctuations is considered. In the tight-binding regime, we employ Wannier functions to derive a discrete nonlinear Schr\"odinger equation that captures the dynamics of the system. 
 Using linear stability analysis coupled to multiple scale expansion, we systematically study the modulational and parametric instabilities of the system, identifying the conditions under which these instabilities emerge. The corresponding instability domains are found and examined. The roles of long-range interactions and quantum fluctuations are highlighted, demonstrating their significant impact on the stability and dynamics of the lattice system. Our analytical predictions are validated through direct numerical simulations, which confirm the instability diagrams through the effective onset of instabilities and reveal the intricate interplay between interaction strength, quantum fluctuation, and the long-range nature of the dipole-dipole forces. This work provides insights into the control and manipulation of ultracold quantum gases in optical lattices, which have potential applications in quantum simulation and condensed matter physics.

%We consider a dilute and ultracold bosonic gas of weakly-interacting atoms with a nonzero interaction range. Using the Bogoliubov-Popov-Beliaev approximation, we derive a modified Gross–Pitaevskii equation with beyond-mean-field corrections due to quantum depletion and anomalous density. This result is obtained from the stationary equation of the Bose–Einstein order parameter coupled to the Bogoliubov–de Gennes equations of the out-of-condensate field operator. We show that, in the presence of a generic external trapping potential, the key steps to get the modified Gross–Pitaevskii equation are the semiclassical approximation for the Bogoliubov–de Gennes equations, a slowly-varying order parameter and a small quantum depletion. In the uniform case, from the modified Gross–Pitaevskii equation, we get the familiar equation of state with Lee–Huang–Yang correction.

\end{abstract}

\maketitle

\section{Introduction}

Ultracold quantum gases trapped in optical lattices have emerged as a versatile platform for exploring fundamental phenomena in quantum many-body systems, including superfluidity, quantum phase transitions, and nonlinear dynamics~\cite{Bloch-RevModPhys-80-885-2008,Lewenstein2012,Jaksch-AnnPhys-315-52-2005}. The high degree of controllability in these systems, achieved through tunable lattice depths, interatomic interactions, and external fields, has enabled the simulation of complex condensed matter systems and the study of novel quantum states~\cite{Greiner-Nature-415-39-2002,Bakr-Nature-8462-74-2009,Bloch-NaturePhys-8-267-2012,Georgescu-RevModPhys-86-153-2014}. The broad range of phenomena exhibited by these systems includes parametric and modulational instabilities, which have garnered significant attention due to their role in the breakdown of coherent states and the emergence of complex dynamical behavior~\cite{Kevrekidis2008,Staliunas2003, Kuznetsov-PhysRep-142-103-1986}. These instabilities are particularly relevant in the context of Bose-Einstein condensates (BECs), where nonlinear interactions and lattice periodicity can lead to the amplification of small perturbations, driving the system into unstable regimes~\cite{Wamba-PhysLettA-375-4288-2011, Armaroli-OptExpr-20-25096-2012}. In the context of media described by the nonlinear Schr\"odinger equation and its variants, parametric instability refers to a phenomenon where small perturbations grow exponentially in a system at parametric resonance due to periodic modulation of system parameters~\cite{Trillo-JOptSocAmB-6-889-1989,Rapti-JPhysB-37-S257-2004}. Such a phenomenon is of particular interest as it can lead to the emergence of novel quantum states and dynamical patterns~\cite{Kivshar2003}.

Recent advances in the manipulation of ultracold atoms with long-range interactions, such as those arising from dipole-dipole forces of dipolar atoms and molecules or Rydberg atoms, have opened new avenues for exploring the interplay between nonlocal interactions and lattice effects~\cite{Lahaye-RepProgPhys-72-126401-2009,Baranov-ChemRev-112-5012-2012,Frisch-PhysRevLett-115-203201-2015,Chomaz-RepProgPhys-86-026401-2023,Gallagher-ADvAtomMolOptPhys-56-161-2008}. Unlike short-range contact interactions, long-range interactions introduce additional complexity into the system, leading to novel instability mechanisms and dynamical features~\cite{Goral-PhysRevLett-88-170406-2002, Pupillo-PhysRevLett-104-223002-2010}. In such a context, however, the effect of long-range interactions on parametric instabilities remains less understood. 
Furthermore, the inclusion of quantum fluctuations, often neglected in mean-field descriptions, has been shown to play a crucial role in stabilizing or destabilizing quantum gases, particularly in regimes where nonlinear effects dominate~\cite{Petrov-PhysRevLett-115-155302-2015, Wachtler-PhysRevA-93-061603-2016}.
For instance, zero-point quantum fluctuations, which are known to stabilize attractive Bose gases against collapse, induce quantum droplets (QDs). These fluctuations are well described theoretically by the Lee–Huang–Yang (LHY) corrections~\cite{Lee-PhysRev-106-1135-1957}. In one~\cite{Astrakharchik-PhysRevA-98-013631-2018}, two~\cite{Petrov-PhysRevLett-117-100401-2016} and three~\cite{Petrov-PhysRevLett-115-155302-2015} dimensions, respectively, the LHY correction in the energy density is proportional to $n^{3/2}$, $n^2 \ln (n)$, and $n^{5/2}$, where $n$ is the condensate density,
which plays an important role in stabilizing QDs. In a Gross-Pitaevski equation, known to well describe the dynamics of quantum gases at ultralow temperatures, the LHY term represents a higher-order nonlinear repulsive interaction, which eliminates the collapse of the gas induced by the mean-field attractive force~\cite{Petrov-PhysRevLett-115-155302-2015,Zheng-FrontPhys-16-22501-2021,Tabi-PhysLettA-485-129087-2023}.

In this work, we intend to investigate the parametric and modulational instabilities of binary ultracold quantum gases with long-range dipole-dipole interactions and quantum fluctuations in deep optical lattices.
We expect the inclusion of quantum fluctuations in our model to allow us to explore regimes where mean-field theory breaks down, and provide a more comprehensive understanding of the instability mechanisms in ultracold quantum gases. 
The organization of the work is as follows: 
In Section~\ref{sec:model}, starting from an extended Gross-Pitaevskii (eGP) model that incorporates long-range interactions and beyond-mean-field corrections, we derive a discrete nonlinear Schr\"odinger equation using Wannier functions, which accurately describe the tight-binding regime of deep optical lattices. In Section~\ref{sec:analytics}, we then perform analytical investigations based on linear stability analysis to identify the conditions under which parametric and modulational instabilities arise. That allows us to emphasize the role of long-range interactions in modifying the stability thresholds and dynamical behavior. In Section~\ref{sec:numerics}, our analytical predictions are validated through direct numerical simulations, which reveal the intricate interplay between lattice depth, interaction strength, and the nonlocal nature of dipole-dipole forces. The work ends in Section~\ref{sec:conclude} with a conclusion along with a summary of our findings.

\section{Model}\label{sec:model}

Consider a binary mixture of Bose-Einstein condensates with dipole-dipole interaction close to the absolute zero. The system is trapped in an optical lattice (OL) given by the (external) potential 
\begin{align}
	V_{\text{ext}}=V_{0x} \sin^2(k x) + V_{0y} \sin^2(k y) + V_{0z} \sin^2(k z) ,
\end{align}
where $k = 2\pi/\lambda$, with $\lambda$ being the laser wavelength, i.e. the lattice spacing is $\lambda/2$. 
The depths $V_{0 i}$ of the lattice in the three directions $i=x,y,z$ are determined by the intensity of the corresponding pair of laser beams, which is easily tunable in an experiment.
The mean-field evolution of the  two-component BEC system is described by the macroscopic wave functions $\Psi_\nu(\mathbf{r},t)$, with $\nu=1,2$. These wave functions obey the following nonlocal Gross-Pitaevskii equations~\cite{Saito-PhysRevLett-102-230403-2009,Xi-PhysRevA-97-023625-2018}:
\begin{align}
	\begin{split}
	\mathrm{i}\hbar\frac{\partial \Psi_\nu}{\partial t}= &\Bigg[ -\frac{\hbar^2 \nabla^2}{2 m_\nu}  +V_{\text{ext}}(\mathbf{r}) + \sum_{\nu'=1}^{2} \mathcal{G}_{\nu \nu'}|\Psi_{\nu'}(\mathbf{r},t)|^2 \\
	 + & \sum_{\nu'=1}^{2} \int \mathcal{U}_{\nu\nu'}(\mathbf{r}-\mathbf{r}') |\Psi_{\nu'}(\mathbf{r}',t)|^2 d^3\mathbf{r}'  
  \Bigg]\Psi_\nu .
\end{split}
\end{align}
When all dipoles are polarized along the $z$ axis, the potential of the dipole-dipole interaction (DDI) can be expressed in the form~\cite{Edmonds-JPhysCommun-4-125008-2020,Chomaz-RepProgPhys-86-026401-2023}
\begin{align}
	\mathcal{U}_{\nu\nu'}(\mathbf{r}-\mathbf{r}')=C_{\nu\nu'}\frac{1-3\cos^2\Theta_{\nu\nu'}}{|\mathbf{r}-\mathbf{r}'|^3} ,
\end{align} % https://ar5iv.labs.arxiv.org/html/0905.0386
where $\Theta_{\nu\nu'}$ is the angle between the relative position vector $\mathbf{r}-\mathbf{r}'$  of the particles and the polarization axis $z$. The prefactor $C_{\nu\nu'}$ is the coupling constant, which depends on the dipole moment $d$ as well as the permeability $\mu_0$ and the permittivity $\varepsilon_0$ of vacuum.
To simplify, we consider a one-dimensional optical lattice described by $V_{\text{ext}}(x)=V_0 \sin^2(k x)$, where the energy barrier $V_0$ between adjacent sites is usually expressed in units of the recoil energy $E_{\text{rec}}=\hbar^2 k^2/(2m)$. Furthermore, we consider a symmetric system, which can consist of equal-mass (or homonuclear) bosonic species with same density and intraspecies scattering length, i.e. the repulsion strengths $g_{11}=g_{22}\equiv g$~\cite{Petrov-PhysRevLett-117-100401-2016,Astrakharchik-PhysRevA-98-013631-2018}. 
Including a LHY contribution due to quantum fluctuations to the mean-field term, and assuming $\Psi_1=\Psi_2\equiv \Psi/\sqrt{2}$, the beyond-mean-field dynamics of the  two-component dipolar BEC system is described by the following extended Gross-Pitaevskii equations~\cite{Astrakharchik-PhysRevA-98-013631-2018,Boettcher-PhysRevRes-1-033088-2019,Zheng-FrontPhys-16-22501-2021,Yang-Photonics-10-405-2023}:
\begin{align}
	\begin{split}
		\mathrm{i}\hbar\frac{\partial \Psi}{\partial t}= &\Bigg[ -\frac{\hbar^2}{2 m}\frac{\partial^2 }{\partial x^2}   + \delta{\mathcal{G}}|\Psi|^2 -\frac{\sqrt{2m}}{\pi \hbar} \mathcal{G}^{3/2}|\Psi| \\
		+ & V_{\text{ext}}(x) + \int \mathcal{U}(x-x') |\Psi(x',t)|^2 dx'  
		\Bigg]\Psi ,
	\end{split}
\end{align}
where parameters $\mathcal{G}$ and $\delta{\mathcal{G}}\equiv \mathcal{G}+\mathcal{G}_{12}$ are positive. For the sake of generality, one can make the interaction strength to be time-dependent, i.e.  $\mathcal{G} \equiv \mathcal{G}(t)$. That is experimentally achieved by inducing Feshbach resonances~\cite{Chin-RevModPhys-82-1225-2010,Chomaz-RepProgPhys-86-026401-2023}, which are particularly convenient tools in ultracold atomic systems for changing the scattering length. In particular, one can make a periodic driving of the interaction strength, which may allow us, for instance, to achieve parametric resonances.

For deep OLs, the interwell barriers of the periodic structure are high enough, i.e. $V_0\gg \mu$, where $\mu$ is the chemical potential. Then, the wells, which are centered at the minima of the laser potential $V_{\text{ext}}$ located at points $x_n = n \lambda/2$, with $n$ integer, can be assumed to be enough isolated from each other. Such a system can be studied using the tight binding model, which substitutes each wave function by an approximate set of wave functions $\psi_{n}(t)$ at different lattice sites $n$. The wave functions are separated into a complex dynamical part $\psi_{n}$ and real spatial part $\Phi_n$ through the nonlinear tight binding ansatz~\cite{Jaksch-AnnPhys-315-52-2005,Smerzi-PhysRevA-68-023613-2003,Trombettoni-JPhysB-39-S231-2006}
\begin{align}
	\Psi(x,t)=\sum_n \psi_{n}(t) \Phi_n(x-x_n;N_n(t)) ,
\end{align}
where $N_{n}(t)\equiv |\psi_{n}|^2$ is the number of atoms in the $n$th site at time $t$. The total number of atoms of each species in the lattice, $N\equiv \sum_n |\psi_{n}|^2$, ensures the normalization.
The Wannier functions $\Phi_n$, localized around the center of the $n$th well, are normalized to one and quasi-orthogonal, i.e. $\int \Phi_n \Phi_{n\pm 1} dx \simeq 0$ and  $\int |\Phi_n |^2 dx = 1$. Due to deep OLs, only tunnellings to nearest neighbors are important, and terms proportional to the integrals $\int \Phi_n^3 \Phi_{n\pm 1} dx $ and $\int \Phi_n^2 \Phi_{n\pm 1}^2 dx $  are neglected for the sake of simplicity. Note in passing that for deep enough lattices, the Wannier functions are well approximated by Gaussians~\cite{Natale-CommunPhys-5-227-2022}. Then, the discrete nonlinear Schr\"odinger equation (DNLS) for the set of macroscopic wavefunctions $\psi_n(t)$ can be derived by multiplying the equation on the left by $\Phi_n^*$ and integrating over space. We obtain the following set of effective one-dimensional
DNLS equations:
%
% \begin{widetext}
%
% \end{widetext}
%
\begin{equation}\label{eq:model1}
	\begin{split}
	\mathrm{i}\hbar\frac{\partial\psi_{n}}{\partial t} = &-J(\psi_{n+1}+\psi_{n-1}) +U_c(t)|\psi_{n}|^{2}\psi_{n} +\mathcal{E}_n\psi_{n} \\
	+ & U_d(|\psi_{n+1}|^{2}+|\psi_{n-1}|^{2})\psi_{n} -U_{q}(t)|\psi_{n}|\psi_{n} .
\end{split}
\end{equation}
We considered an average number $N_0$ of atoms per site, and kept only the zero-order terms, such that $\Phi_n(x-x_n, N_n)\to\Phi_n(x-x_n, N_0)$. The hopping parameter $J$,  (i.e. the tunneling rate between two neighboring lattice sites), the cubic nonlinear coefficient $U_c$, the effective strengths of DDI $U_d$, and quantum fluctuation $U_q$ are expressed through overlap integrals of the Wannier functions  as follows:
\begin{equation}
	\begin{split}
	J &\simeq -\int dx \, \Bigg[  \frac{\hbar^2}{2 m}  \frac{\partial \Phi_n^*}{\partial x} \frac{\partial \Phi_{n\pm 1}}{\partial x}    + \Phi_n^*  V_{\text{ext}}(x)  \Phi_{n\pm 1} \Bigg], \\
	U_c(t) & \simeq \delta{\mathcal{G}}(t) \int dx \,|\Phi_n|^4 , \ \delta{\mathcal{G}}(t) \equiv \mathcal{G}_{12}+\mathcal{G}(t), \\ 
	\mathcal{E}_n &\simeq \int dx \, \Phi_n^* \Bigg[ -\frac{\hbar^2}{2 m}\frac{\partial^2 }{\partial x^2}   + V_{\text{ext}}(x) \Bigg] \Phi_{n}, \\
	U_d &\simeq \int dx dx' \, |\Phi_n^*(x)|^2 C_{dd}\frac{1-3\cos^2\Theta}{|x-x'|^3} |\Phi_{n\pm 1}(x')|^2 \\ 
	U_q(t) &\simeq \int dx \, \frac{\sqrt{2m}}{\pi \hbar} \mathcal{G}(t)^{3/2}|\Phi_{n}|^3 .
\end{split}
\end{equation}
Let us mention in passing that it is the orientation of dipoles, which determines the sign of $U_d$. For dipolar particles in a head-to-tail configuration, we have $\Theta=0$, i.e. $U_d<0$, and the dipole-dipole interaction is attractive. Meanwhile, for dipoles sitting side by side, we have $\Theta=\pi/2$, i.e. $U_d>0$, and the dipole-dipole interaction is repulsive. The interaction vanishes at the magic angle $\Theta=\arccos(1/\sqrt{3})\approx 54.7^\circ$. When the species have a quite strong dipolar moment, the DDI term in Eq.~\eqref{eq:model1} may even contain next nearest neighbor components and beyond, as shown in Ref.~\cite{Natale-CommunPhys-5-227-2022}. 
The explicit expressions of parameters $J$, $U_c$, $U_d$ and $U_q$ are not important in the present work, but they may be found if one has an approximate expression of the ground-state wave function. To further simplify Eq.~\eqref{eq:model1}, let us introduce the rotational frame change $\psi_{n}\to \psi_{n} \exp(-\mathrm{i}\mathcal{E}_n t/\hbar)$ and $t\to t J$, and work in units where $\hbar=1$. Then, the DNLS equation that rules the system reads 
\begin{equation}\label{eq:modelfinal}
	\begin{split}
		\mathrm{i}\frac{\partial\psi_{n}}{\partial t} = &-(\psi_{n+1}+\psi_{n-1}) +g(t)|\psi_{n}|^{2}\psi_{n}  \\
		& +  \gamma(|\psi_{n+1}|^{2}+|\psi_{n-1}|^{2})\psi_{n} -\eta(t)|\psi_{n}|\psi_{n} ,
	\end{split}
\end{equation}
where $g(t)=U_c(t)/J$, $\gamma=U_d/J$, and $\eta(t)=U_q(t)/J$. For a system under Feshbach resonance management, one may set $\mathcal{G}(t)=\mathcal{G}_c+\mathcal{G}_1\sin(\omega t)$, which implies that the rescaled DDI and LHY coefficients may have the form $g(t)\equiv g_0 [1+\epsilon \sin(\omega t)]$ and $\eta(t) \equiv \eta_0 [1+\epsilon \sigma \sin(\omega t)]$, where  
\begin{equation}
	\begin{split}
		g_0 &=(\mathcal{G}_{12}+\mathcal{G}_c) \int dx \,|\Phi_n|^4, \ \epsilon = \frac{\mathcal{G}_1}{\mathcal{G}_{12}+\mathcal{G}_c}\\ 
		\eta_0 &= \int dx \, \frac{\sqrt{2m}}{\pi \hbar} \mathcal{G}_c^{3/2}|\Phi_{n}|^3, \ \sigma \simeq \frac{3}{2}\left(1+\frac{\mathcal{G}_{12}}{\mathcal{G}_c}\right) .
	\end{split}
\end{equation}
Without loss of generality, we restrict ourself to the particular case when the dimensionless parameter $\sigma=1$, which is achieved for $\mathcal{G}_{12}\simeq -\mathcal{G}_{c}/3$. This is actually not a shortcoming of the model because any arbitrary values of $\mathcal{G}_{12}$ can be experimentally realized. Furthermore, the results for any $\sigma$ can be straightforwardly deduced following the steps of the current analysis.

\section{Analytical investigations}\label{sec:analytics}

While modulational and parametric instabilities are characterized by an exponential growth in the wave amplitude, the latter occurs at best in a modulationally stable system under the effect of periodic driving. Then, we will first study the conditions under which modulated plane waves become unstable in the system ruled by Eq.~\eqref{eq:modelfinal} under a time-independent particle-particle interaction strength. Next, we will consider time-periodic modulation of that interaction strength and examine parametric instability in the system. 

\subsection{Modulational instability}\label{sec:modulinstab}

In other to examine the instability of modulated plane wave under perturbation, we use the linear stability approach, which directly allows getting the instability criteria through a linearization process.

\subsubsection{Linear stability analysis}

The stationary carrier wave solution of the DNLS Eq.~\eqref{eq:modelfinal} is a plane wave of the form
\begin{equation}\label{eq:planewave1}
	\psi_n(t) = u_0 e^{\mathrm{i} \left(q n + 2t\cos q - \int_{0}^{t} \left[ g(s)|u_0|^2 -\eta(s) |u_0| + 2\Delta_1\right] \right)} ,
\end{equation}
where $\Delta_1\equiv \gamma |u_0|^2$. To examine the modulational instability (MI) of the BEC system via the linear stability analysis, we introduce a perturbed ansatz by replacing $u_0$ in Eq.~\eqref{eq:planewave1} by $u_0(1+\delta{\phi}_n(t))$, where $\delta{\phi}_n(t)$ is a small complex perturbation, i.e. $|\delta{\phi}_n|\ll 1$.
Substituting the perturbed plane wave into the DNLS Eq.~\eqref{eq:modelfinal}, and neglecting terms of order higher than one in $\delta{\phi}_n $and its complex conjugate $\delta{\phi}_n^*$, we find that the evolution of the perturbation obeys the linearized equation:
\begin{equation}\label{eq:perturbwave1}
	\begin{split}
	\mathrm{i} \frac{\partial(\delta{\phi}_n)}{\partial t} = -\left( \delta{\phi}_{n-1} e^{-\mathrm{i}q} +\delta{\phi}_{n+1} e^{\mathrm{i}q} -2\cos q \, \delta{\phi}_n \right)  \\
	+\Delta(t)\left( \delta{\phi}_n^* + \delta{\phi}_n\right) \\+ \Delta_1 \left( \delta{\phi}_{n+1}^* +\delta{\phi}_{n-1}^*+ \delta{\phi}_{n+1} +\delta{\phi}_{n+1} \right),
\end{split}
\end{equation}
where $\Delta(t) \equiv g(t)  |u_0|^2 - \eta(t)|u_0|/2$, which can be readily rewritten as $\Delta(t) = \Delta_0 [1+\epsilon \sin(\omega t)]$, with $\Delta_0 \equiv g_0 |u_0|^2 - \eta_0|u_0|/2$. Let us look for plane wave perturbations of the form
\begin{equation}\label{eq:ansatzwave1}
	\delta{\phi}_{n}(t) = \alpha(t) e^{\mathrm{i}k n} + \beta(t)^* e^{-\mathrm{i}k n} ,
\end{equation}
where $k$ is the perturbation wavenumber, $\alpha(t)$ and $\beta(t)$ are time-dependent complex fields, which are weak compared to one. Replacing the ansatz~\eqref{eq:ansatzwave1} into Eq.~\eqref{eq:perturbwave1} and linearizing, we obtain the following pair of first-order coupled equations:
\begin{equation}\label{eq:perturbwave2}
	\begin{split}
		\mathrm{i} \frac{d \alpha}{d t} & =  \left(B+C\right) \, \alpha + \Gamma (\alpha +\beta) ,
		\\
		\mathrm{i} \frac{d \beta}{d t} & = \left(B-C\right) \, \beta - \Gamma (\alpha +\beta),
	\end{split}
\end{equation}
where $B=2\sin q \sin k$, $C=2\cos q \,(1-\cos k)$, and $\Gamma(t) = \Delta(t) +2\Delta_1 \cos k$, which can be rewritten as $\Gamma(t) = \Gamma_0 +\epsilon \Delta_0 \sin(\omega t)$, with $\Gamma_0 \equiv \Delta_0+2\Delta_1 \cos k$ being the effective nonlinearity strength of the system. The above two equations can be combined to form a second-order equation as follows:
\begin{equation}\label{eq:perturbwave3}
	\begin{split}
		\frac{d^2 \alpha}{d t^2}  =  \bigg[2\mathrm{i}B &-\frac{d(\ln \Gamma)}{d t}\bigg]\frac{d \alpha}{d t} +\bigg[C^2-B^2+2C\Gamma\\ &-\mathrm{i}(C+B)\frac{d(\ln \Gamma)}{d t}\bigg] \alpha =0 .
	\end{split}
\end{equation}
\begin{figure}[t!]
	\centering	
	\includegraphics[width=0.50\textwidth]{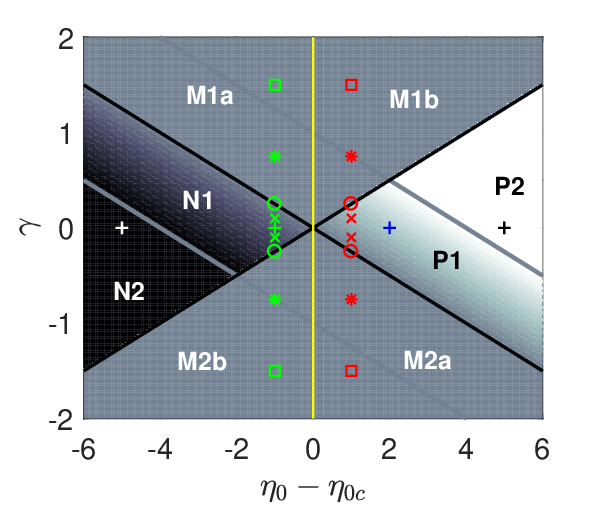}
	\caption{Parameter subdomains that determine the stability of the system in the $(\eta_0,\gamma)$-plane. We have $u_0=1.0$ and $g_0=1.0$, which implies $\eta_{0c}=2$. Markers depict some selected points where the structure of gain profiles will be examined.}\label{fig:InstabDigram1}
\end{figure}
\begin{figure}[t!]
	\centering	
	\includegraphics[width=0.23\textwidth]{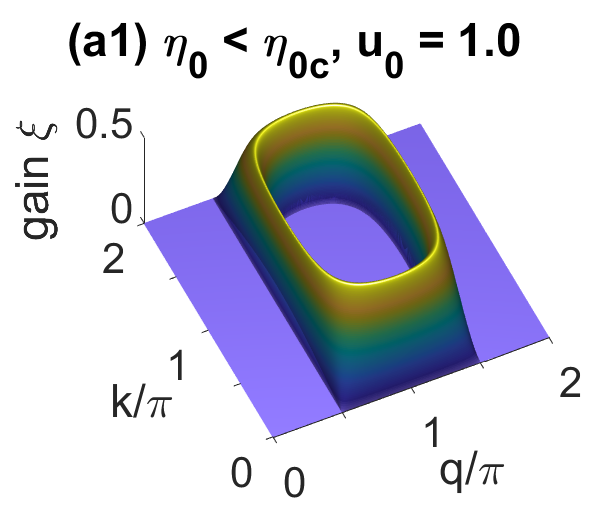}
	\includegraphics[width=0.23\textwidth]{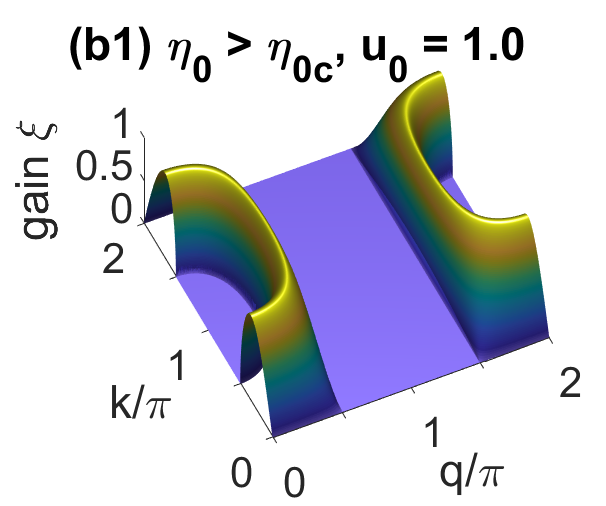}
	\includegraphics[width=0.23\textwidth]{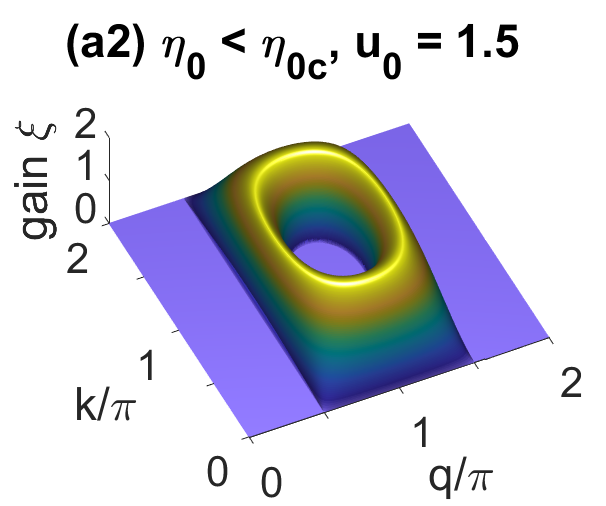}
	\includegraphics[width=0.23\textwidth]{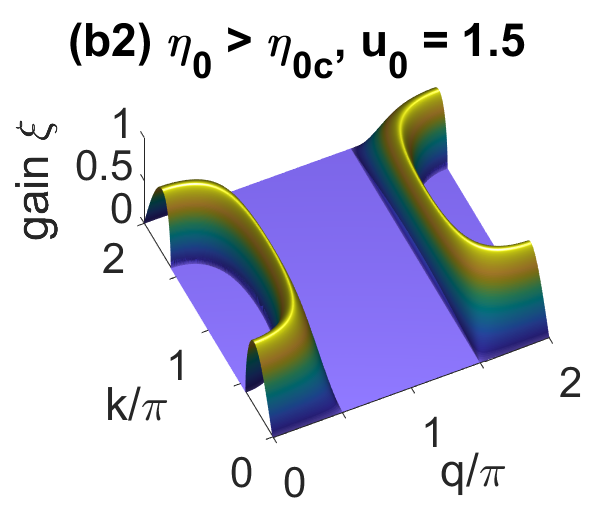}
	\includegraphics[width=0.23\textwidth]{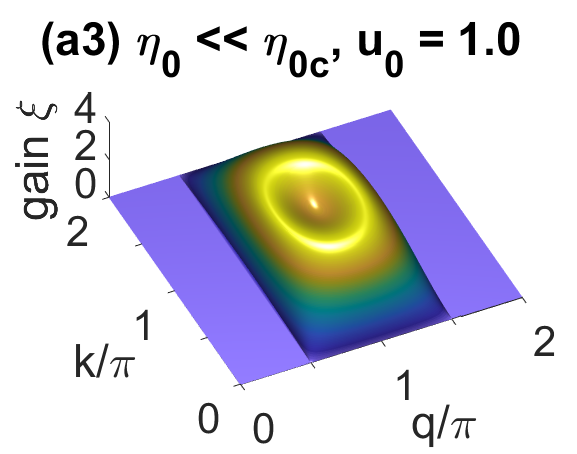}
	\includegraphics[width=0.23\textwidth]{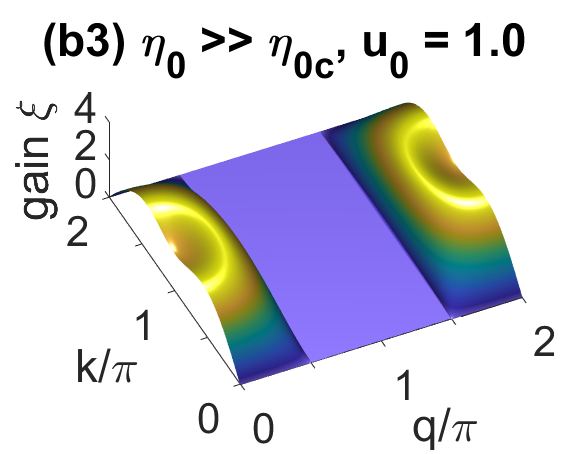}
	\caption{Effect of wave amplitude $|u_0|$ on the modulational instability gain profiles analytically predicted for (a1) $\eta_0=1$, $u_0=1.0$; (b1) $\eta_0=4$, $u_0=1.0$; (a2) $\eta_0=1$, $u_0=1.5$; (b2) $\eta_0=4$, $u_0=1.5$; (a3) $\eta_0=-3$, $u_0=1.0$; (b3) $\eta_0=7$, $u_0=1.0$. The other parameters are $\gamma=0$, $N=2048$, $g_0=1.0$. For $\eta_0=2g_0 |u_0|\equiv \eta_{0c}$, the gain is zero at all $(q,k)$ points. We have $\eta_{0c}=2$ and $3$ for $u_0=1.0$ and $1.5$, respectively. Panels (a3), (a1), (b1) and (b3) correspond to white, green, blue and black plus ($+$) markers, respectively, depicted in Fig.~\ref{fig:InstabDigram1}.}\label{fig:MIgainNoDDI1}
\end{figure}
In the static case when $\epsilon=0$, Eq.~\eqref{eq:perturbwave3} corresponds to a second-order equation with constant coefficients, whose solution $\alpha(t) \propto \exp\left(\mathrm{i}B\pm \sqrt{-(C^2+2C\Gamma_0)} t\right)$, which yields the MI condition
\begin{align}\label{eq:MIcond1}
	C(C+2\Gamma_0)\equiv \Omega^2 <0.
\end{align}
In that case, the local growth rate of instability corresponding to unstable modes is given by  
$\xi = \mathrm{Im}\left(\Omega\right)$.
Obviously the MI condition~\eqref{eq:MIcond1} is in accordance with the results of the BECs with two-body interatomic
interaction~\cite{Rapti-JPhysB-37-S257-2004}. Indeed, in the absence of both quantum fluctuations (QFs) and dipole-dipole interaction, i.e. $\eta_0=\gamma=0$, we have $\Gamma_0=g_0 |u_0|^2$, and retrieve the MI criteria 
\begin{align}\label{eq:MIcond2}
	\cos^2 q \,(1-\cos k)+g_0 |u_0|^2 \cos q <0,
\end{align}
for plane waves of background wave number $q$ in a standard BEC with contact two-body interaction embedded  in a deep OL~\cite{Trombettoni-JPhysB-39-S231-2006} and perturbed with wave numbers $k$. It is worth mentioning that the instability criterion~\eqref{eq:MIcond1} is symmetric in the $(q,k)$ domain, since it is invariant under the change $q\to -q$ and $k\to -k$. Therefore it is enough to only focus on positive $q$ and $k$.
Out of all wavenumbers that satisfy the instability condition~\eqref{eq:MIcond1}, there exists particular ones, which yield the largest instability growth rate in the system. In safe parameter domains, these critical wavenumbers are given by
\begin{align}\label{eq:kmax}
	k_{\text{cr}} =\left\{ \begin{tabular}{l l}
	$\pm \arccos\left(\frac{\Delta_0-2\Delta_1+2\cos q}{-4\Delta_1+2\cos q}\right)$,	& $q\in ]-\frac{\pi}{2}, \frac{\pi}{2} [$\\
	$\pi$, & \ $q\in ]\frac{\pi}{2}, \frac{3\pi}{2}[ $,
	\end{tabular} \right.
\end{align}
where $k$ and $q$ are defined modulo $2\pi$. The corresponding maximum instability growth rate, $\xi_{\text{max}} \equiv \xi(k=k_{\text{cr}}) $, is given by
\begin{align}\label{eq:Gainmax}
	\begin{split}
\xi_{\text{max}} =&\mathrm{Re}\Bigg( \bigg[\frac{(\Delta_0+2\Delta_1)^2\cos q}{-2\Delta_1+\cos q}\bigg]^{1/2} \\
& + 2\sqrt{2} \bigg[ -\cos q (\Delta_0-2\Delta_1+2\cos q) \bigg]^{1/2} \Bigg) .
\end{split}
\end{align}
Compared to the gain $\xi$ per wavenumber $k$, the maximum gain $\xi_{\text{max}}$ has the advantage to better reproduce the results of computer experiments with single-wavelength or noisy perturbations and even laboratory experiments.

\begin{figure}[t!]
	\centering	
	\includegraphics[width=0.23\textwidth]{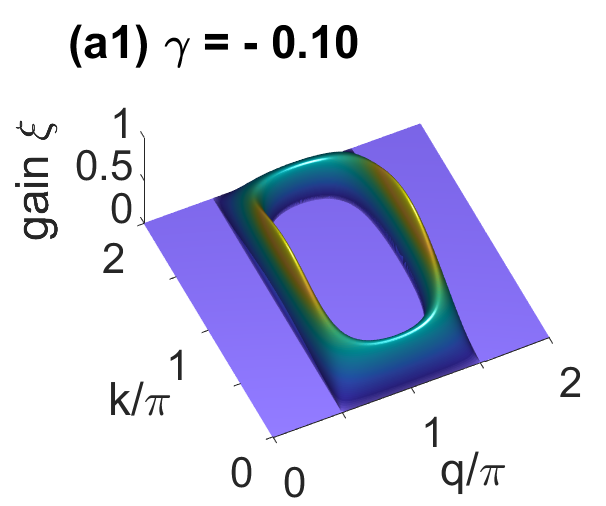}
	\includegraphics[width=0.23\textwidth]{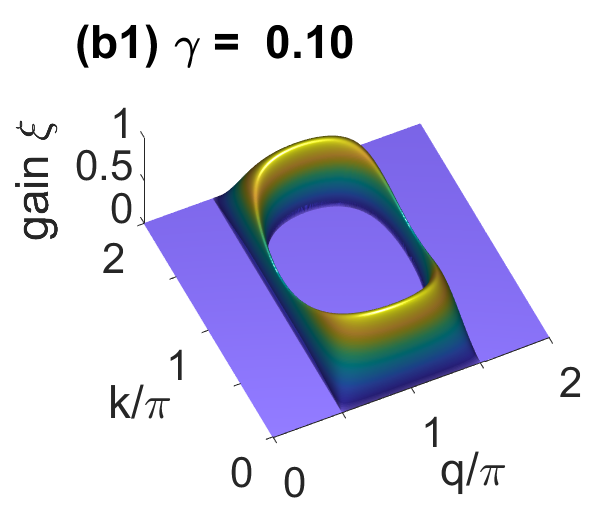}
	\includegraphics[width=0.23\textwidth]{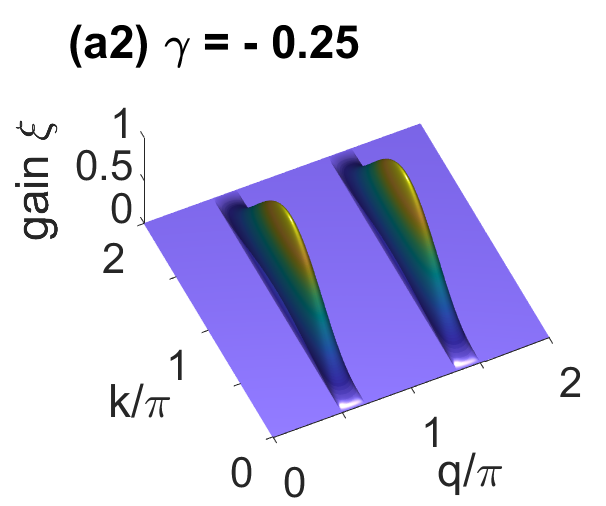}
	\includegraphics[width=0.23\textwidth]{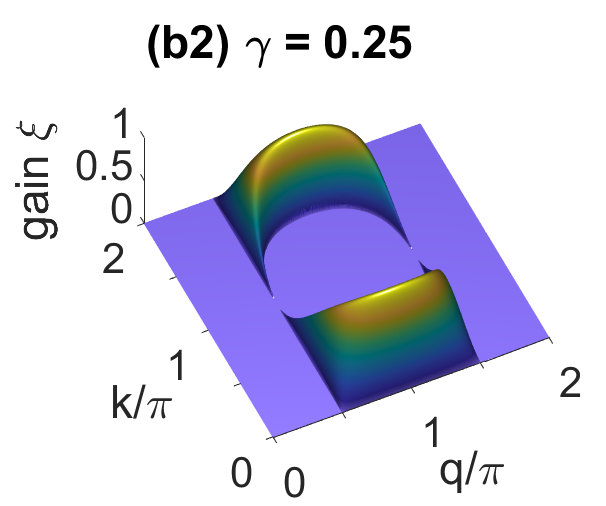}
	\includegraphics[width=0.23\textwidth]{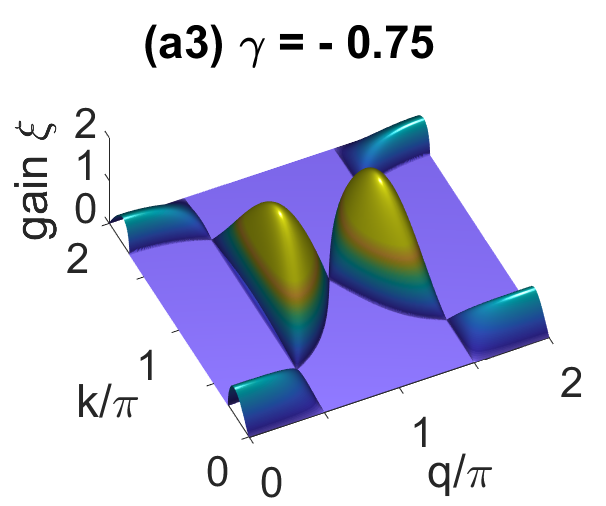}
	\includegraphics[width=0.23\textwidth]{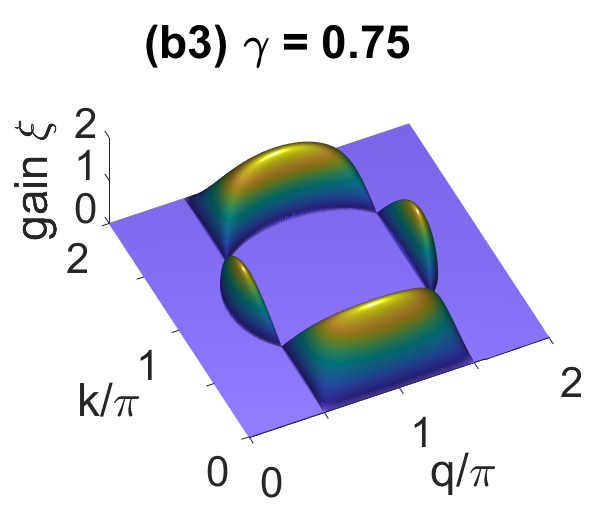}
	\includegraphics[width=0.23\textwidth]{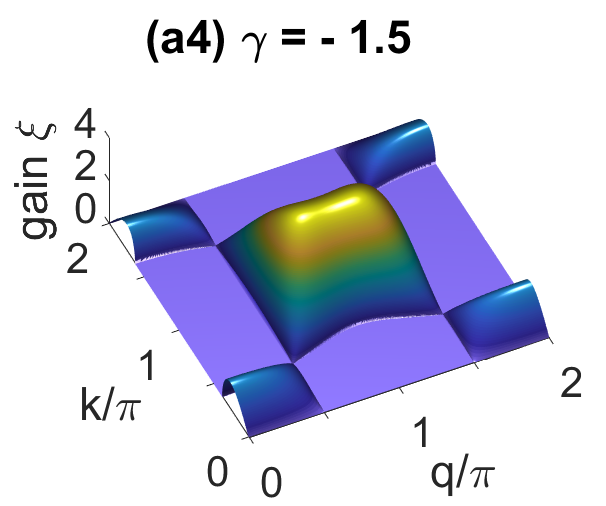}
	\includegraphics[width=0.23\textwidth]{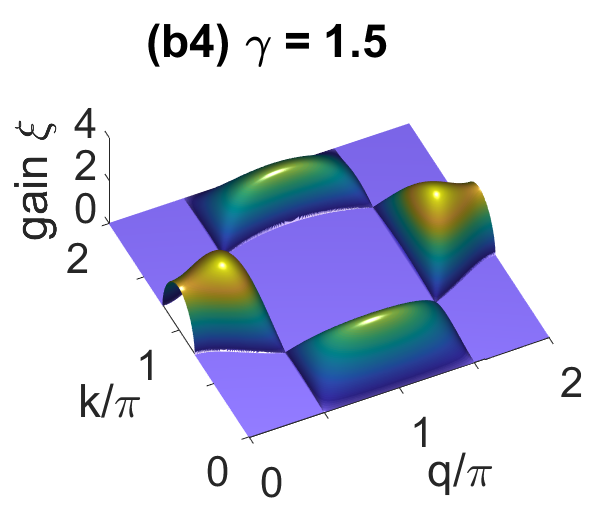}
	\caption{Effect of QFs strength $\gamma$ on the modulational instability gain profiles analytically predicted for (a) $\gamma<0$ (attractive DDI), (b) $\gamma>0$ (repulsive DDI) on the BEC side ($\eta_0<\eta_{0c}$). We used $\eta_0=1$, $N=2048$, $u_0=1.0$, $g_0=1.0$. The case $\gamma=0$ was already presented in Fig.~\ref{fig:MIgainNoDDI1}(a1), with same parameters. Panels (a1)-(b1), (a2)-(b2), (a3)-(b3), and (a4)-(b4) correspond to green cross, circle, asterisk and square markers, respectively, depicted in Fig.~\ref{fig:InstabDigram1}.}\label{fig:MIgainwithDDI1}
\end{figure}

\begin{figure}[t!]
	\centering	
	\includegraphics[width=0.23\textwidth]{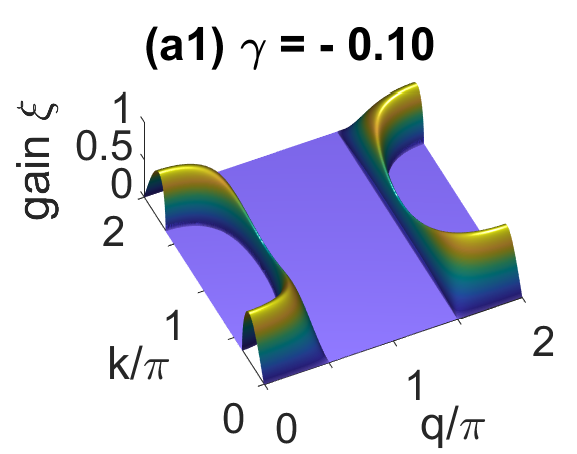}
	\includegraphics[width=0.23\textwidth]{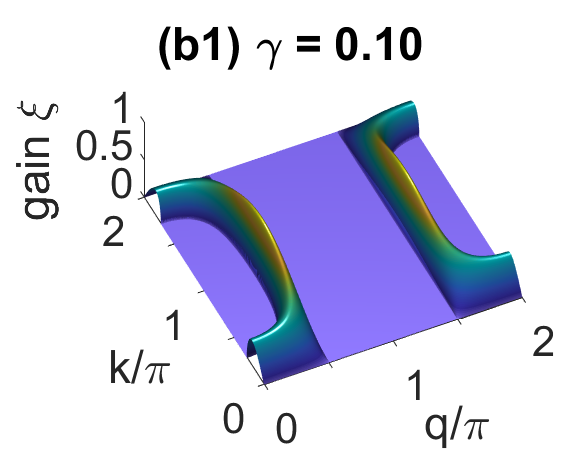}
	\includegraphics[width=0.23\textwidth]{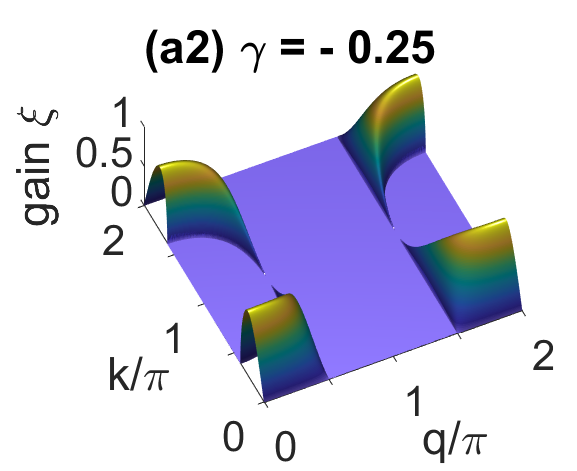}
	\includegraphics[width=0.23\textwidth]{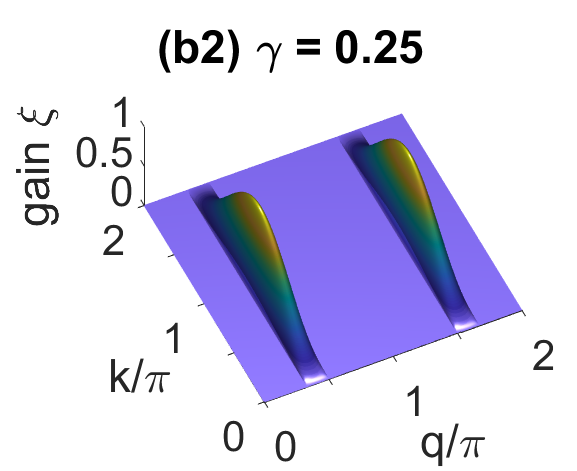}
	\includegraphics[width=0.23\textwidth]{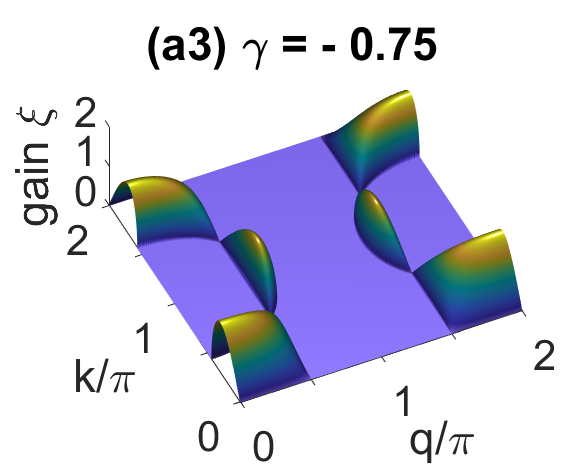}
	\includegraphics[width=0.23\textwidth]{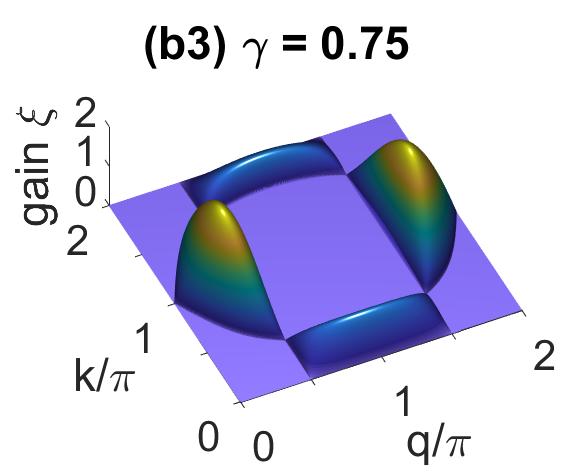}
	\includegraphics[width=0.23\textwidth]{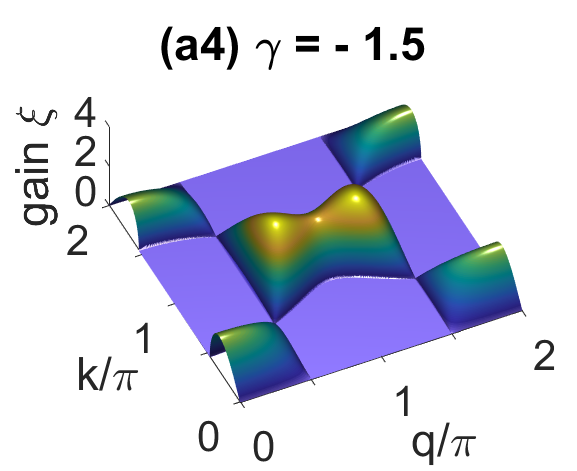}
	\includegraphics[width=0.23\textwidth]{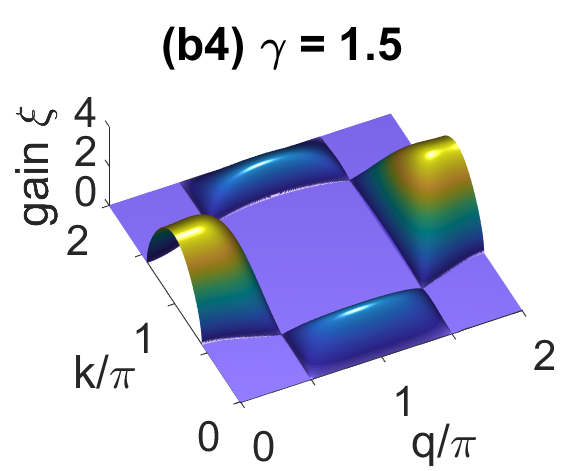}
	\caption{Effect of QFs strength $\gamma$ on the modulational instability gain profiles analytically predicted for (a) $\gamma<0$ (attractive DDI), (b) $\gamma>0$ (repulsive DDI) on the QDs side ($\eta_0>\eta_{0c}$). We used $\eta_0=3$, $N=2048$, $u_0=1.0$, $g_0=1.0$. The case $\gamma=0$ was already presented in Fig.~\ref{fig:MIgainNoDDI1}(a1), with same parameters. Panels (a1)-(b1), (a2)-(b2), (a3)-(b3), and (a4)-(b4) correspond to red cross, circle, asterisk and square markers, respectively, depicted in Fig.~\ref{fig:InstabDigram1}.}\label{fig:MIgainwithDDI2}
\end{figure}

\subsubsection{Instability domains}  

From the obtained instability criteria, the strength of quantum fluctuations may allow us to distinguish two main regimes: the BEC and quantum droplet regimes, given by $\eta_0<\eta_{0c}$ and $\eta_0>\eta_{0c}$, respectively, with $\eta_{0c}=2g_0|u_0|$. The left (BEC) and right (QDs) sides of Fig.~\ref{fig:InstabDigram1}, separated by a red vertical line, depict these regimes. They were obained by looking at the behavior of the term $\Gamma_0/(1-\cos k)$ deductible from the condition~\eqref{eq:MIcond1}. The whole set of subdomains presents a $180^\circ$ rotation symmetry around the origin $(\eta_0-\eta_{0c},\gamma)=(0,0)$. The subdomains are demarcated by well-known straight lines. From top to bottom, the three lines with negative slope are given by  $\gamma=-\frac{1}{4 |u_0|}(\eta_0-\eta_{0c})+\frac{1}{|u_0|^2}$, $\gamma=-\frac{1}{4|u_0|}(\eta_0-\eta_{0c})$, and $\gamma=-\frac{1}{4|u_0|}(\eta_0-\eta_{0c})-\frac{1}{|u_0|^2}$. The line with positive slope is given by  $\gamma=\frac{1}{4|u_0|}(\eta_0-\eta_{0c})$.
Each side contains four subdomains in which the system behaves differently. Considering the wave numbers, one may discriminate the intervals of interest $q\in ]-\pi/2, \pi/2[$ (long wavelengths) and $q\in ]\pi/2, 3\pi/2[$ (short wavelengths) $\mathrm{mod}[2\pi]$.
(i) In subdomains N, for any $q\in ]-\pi/2, \pi/2[$, there exists no value of the perturbation wave number $k$, which can yield the instability in the system. Modes with $q\in ]\pi/2, 3\pi/2[$ are expected to be unstable in subdomain N1 for only some values of $k$, and unstable in subdomain N2 for all values $k$.
(ii) In subdomains P, all modes with $q\in ]\pi/2, 3\pi/2[$ are stable for any perturbation wave number $k$. Meanwhile, modes with $q\in ]-\pi/2, \pi/2[$ are expected to be unstable in subdomain P1 for only some values of $k$, and unstable in subdomain P2 for all values $k$.
(iii) In subdomains M, i.e. M$_{\text{1a}}$, M$_{\text{1b}}$, M$_{\text{2a}}$ and M$_{\text{2b}}$, for any given value of $q$, they will always exist at least one value of $k$ such that the mode $(q,k)$ is stable, and at least one value of $k$ such that the mode is unstable.

\subsubsection{Instability gain profiles} 

To display the instability gain profiles corresponding to the above instability domains, we first consider the system in the absence of dipole-dipole interaction, i.e. $\gamma=0$. The MI gain profiles in the $(q,k)$ plane are shown in Fig.~\ref{fig:MIgainNoDDI1}, where panels (a1) and (a2) correspond to the BEC side while (b1) and (b2) correspond to the QDs side. We realize that, for the QDs and BEC regimes, respectively, the unstable bandwiths are located in the long wavelengths interval $q\in ]-\pi/2, \pi/2[$ and short wavelengths interval $q\in ]\pi/2, 3\pi/2[$. A disk-shaped stability window exists at the center of each unstable region. Furthermore, in the BEC regime, as the density increases, the instability gain increases while the domain of unstable wave numbers in the $(q,k)$ plane broadens. In the QDs regime, the system behaves differently, as the unstable bandwiths become narrower and the MI gain smaller when the density increases. It is worth mentioning that away from the critical value $\eta_{0c}$, the QF strength enhances the instability in both regimes, both by enlarging the bandwidths and growth rate of instability. For strengths $\eta_0$ such that $|\eta_0-\eta_{0c}|\leqslant 4$, as portrayed in panels (a3) and (b3), the stability window closes and all modes $(q,k)$ in the respective regimes and $q$-interval become unstable. 
Besides, unconditional MI suppression, for all perturbation wave numbers $k$, is expected for long and short wavelengths, respectively, in the BEC and QDs regimes, as seen through the zero gain areas in Fig.~\ref{fig:MIgainNoDDI1}.

Let us now consider the system in the presence of dipole-dipole interaction. For a fixed $\gamma\neq 0$, the above described behaviors of the system at $\gamma=0$ still apply to pure BEC and QDs regimes defined by $\eta_0<\eta_{0c-}$ and $\eta_0>\eta_{0c+}$, respectively, where $\eta_{0c\pm}=\eta_{0c} \mp 4\gamma|u_0|$. For strengths of $\eta_0 \in [\eta_{0c-}, \eta_{0c+}]$, however, the system lies in a mixed regime where its stability is conditional, and determined by the specific values of wave numbers. 
For different values of $\gamma$, however, there can be more possible scenarios as displayed in Figs.~\ref{fig:MIgainwithDDI1} and~\ref{fig:MIgainwithDDI2}, on the BEC and QDs sides, respectively, and for both attractive ($\gamma<0$) and repulsive ($\gamma>0$) dipole-dipole interaction. For weak strengths of the DDI, i.e. $|\gamma|\leqslant 0.25 \equiv |\eta_{0c}-\eta_0|/(4 |u_0|)$, corresponding to subdomains N1 and P1 of Fig.~\ref{fig:InstabDigram1}, the there is absence of MI in the long wavelengths interval $q\in ]-\pi/2, \pi/2[$ and short wavelengths interval $q\in ]\pi/2, 3\pi/2[$ on the BEC and QDs sides, respectively. For stronger strengths of the DDI, i.e. $|\gamma|\geqslant 0.25$, corresponding to subdomains M of Fig.~\ref{fig:InstabDigram1}, the maximum value of the gain is higher compared to the previous case, and the absence or presence of MI at both long and short wavelengths depends only the selected pair $(q,k)$. This increase of maximum growth rate as well as the ubiquity of unstable modes at all wavelengths infer a possible enhancement of the instability at large DDI strengths.

These rich results are obtained for $\epsilon=0$, i.e. in the absence of any temporal modulation in the parameters of the system. A natural question that emerges is how such a stability would be impacted by the modulation of contact interaction. We explore this question in the following subsection.

\subsection{Parametric instability}\label{sec:paraminstab}

Similar to modulational instability, we consider an instability, referred to as \textit{parametric}, characterized by an exponential growth in the wave amplitude, happening in the context of periodically driven interactions. 
Up to order $\mathcal{O}(\epsilon)$, the above equation~\eqref{eq:perturbwave3} can be recast as
\begin{align}\label{eq:perturbwave4}
	\begin{split}
		\ddot{\alpha} & =  \left[2\mathrm{i}B-\epsilon \omega\frac{\Delta_0}{\Gamma_0}\cos(\omega t)\right]\dot{\alpha} +\bigg(C^2-B^2+2C\Gamma_0\\ &+\epsilon \bigg[2C \Delta_0 \sin(\omega t) -\mathrm{i}\omega(C+B) \frac{\Delta_0}{\Gamma_0}\cos(\omega t)\bigg] \bigg) \alpha =0 ,
	\end{split}
\end{align}
where the dots stand for the differentiation with respect to time. Because \eqref{eq:perturbwave4} is an ordinary differential equation with periodic coefficients, it is natural to examine whether such a temporal modulation may induce parametric instabilities in the system.

\subsubsection{Instability criteria and domains}
To study parametric instabilities, we consider a \textit{modulationally stable} system, i.e. $\Omega^2>0$, and introduce a regular perturbation expansion of the perturbation amplitude $\alpha$ as a series in $\epsilon$ as follows:
\begin{align}\label{eq:seriesexpand}
\alpha(t)=\alpha_0(t)+\epsilon \alpha_1(t) + \mathcal{O}(\epsilon^2) .
\end{align}
The next step consists in using the ansatz~\eqref{eq:seriesexpand} to solve Eq.~\eqref{eq:perturbwave4} at different orders. It follows that $\alpha_0=e^{\mathrm{i}B}\left( c_1 e^{\mathrm{i}\Omega t}+c_2 e^{-\mathrm{i}\Omega t}\right)$, where $c_{1,2}$ are constants. Secular behaviors in $\alpha$ may arise if the denominator of the first-order term $\alpha_1$ turns to zero, or equivalently if at least one right-hand side term of the $\alpha_1$ equation picks up the natural frequency of the system, i.e. $\omega\pm \Omega=\pm \Omega$. In such a case, we expect the amplitude to diverge. That happens when $\Omega= \omega/2 \equiv \Omega_0$, i.e. the driving frequency is twice the natural frequency, and thus
$C=-\Gamma_0 + \sqrt{\Gamma_0^2 +\omega^2/4} \equiv C_0$,
which corresponds to 
\begin{equation}\label{eq:condq}
	 q=\pm \arccos(C_0/C_m)\equiv \pm q_0.
\end{equation}
The parameter $C_m=2(1-\cos k)$ denotes the maximum value of $C$ over the $q$-domain. The condition~\eqref{eq:condq} holds only for driving frequencies $\omega \leqslant 2\sqrt{C_m(C_m+2\Gamma_0)}$, and when the fixed background wave parameters $u_0, k$ and system's parameters $g_0, \gamma, \eta_0$ satisfy the relation $\Gamma_0 \geqslant -C_m/2$.
To explore the dynamics near the first parametric resonance value, we implement a multiple-scale expansion 
by letting $\Omega=\Omega_0 + \Omega_1\epsilon + \mathcal{O}(\epsilon^2) $, with multiple scale ansatz 
\begin{equation}
\alpha(t)=\alpha_0(t,t_1) + \epsilon \alpha_1(t,t_1) + \cdots ,
\end{equation}
where $t_1=\epsilon t$. Since $\Omega$ is a function of $q$, consequently, it follows the expansion $q=q_0 + q_1\epsilon + \mathcal{O}(\epsilon^2) $. Performing the multiple-scale expansion yields
\begin{equation}\label{eq:solq1}
	q_1 = \pm \frac{-\Gamma_0 + \sqrt{\Gamma_0^2 +\frac{\omega^2}{4}}}{4\sqrt{\Gamma_0^2 +\frac{\omega^2}{4}} \, (1-\cos k)\sin q_0} .
\end{equation}
Substituting~\eqref{eq:solq1} into the expansion of $q$, we identify the instability domain boundaries in the $(q,\epsilon)$ domain to be
\begin{equation}\label{eq:paramInstCrit}
	 \epsilon = \pm \frac{2(C_0+\Gamma_0)\sqrt{C_m^2-C_0^2}}{C_0} \, (q-q_0) .
\end{equation}
These instability boundaries may allow us to locate  and draw the parametric instability bands within the modulationally stable parts of the parameter domains.

\begin{figure}[hbt]
	\centering	
	\includegraphics[width=0.50\textwidth]{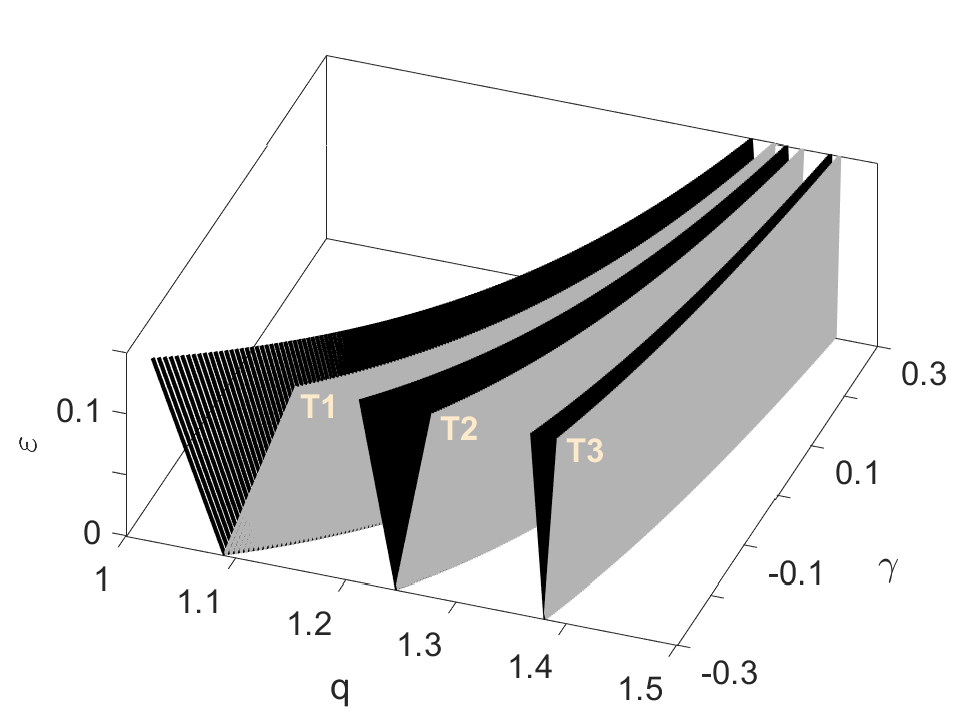}
	\caption{Parametric instability canals in the $(q,\gamma)$-plane at wave numbers around $q=q_0$ for (T1) $\eta_0=-0.5$, (T2) $\eta_0=0$, and (T3) $\eta_0=1.0$. We used $k=0.25\pi$, $\omega=1.0$, $g_0=1$, and $u_0=1$.}\label{fig:paramdom}
\end{figure}

\begin{figure}[hbt]
	\centering	
	\includegraphics[width=0.50\textwidth]{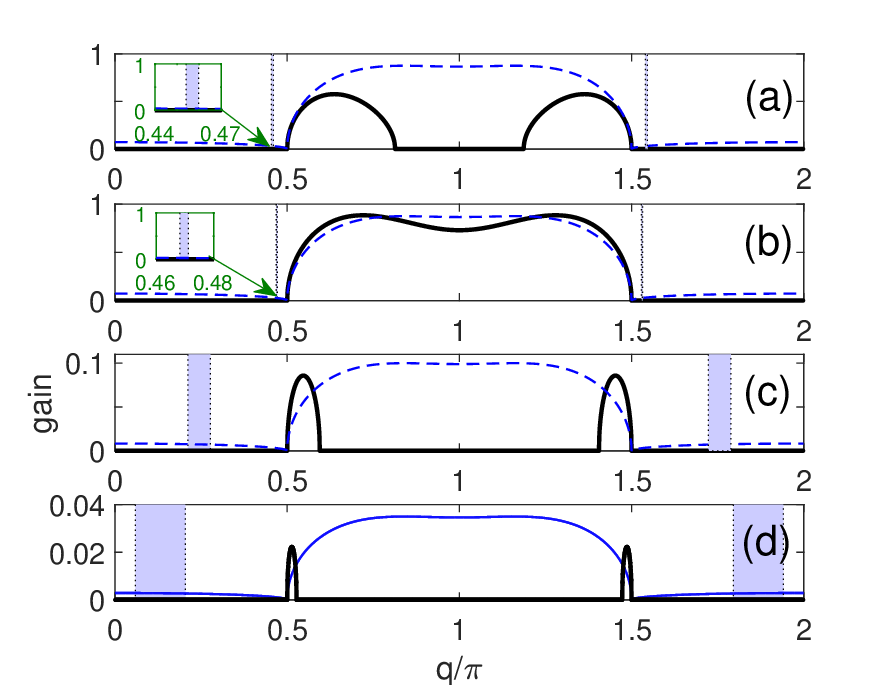}
		\caption{Locations of parametric instability windows within the modulationally stable regions in the $q$-domain, when (a) $k=0.40\pi$, $\gamma=-1.5$; (b) $k=0.40\pi$, $\gamma=-1.0$; (c) $k=0.25\pi$, $\gamma=-1.0$; and (d) $k=0.235\pi$, $\gamma=-1.0$. The other parameters are $\epsilon=0.10$, $\eta_0=-1$, $\omega=1.0$, $g_0=1$, and $u_0=1.0$. Solid black lines depict the instability gain $\xi$ for a specific $k$. Blue dashed lines represent the maximum gain $\xi_{\text{max}}/\ell$, where the scaling factor $\ell=4$, $4$, $35$, and $100$ for (a), (b), (c), and (d), respectively.}\label{fig:paramdom2}
\end{figure}

%\paragraph{Instability domains. }
%
In Fig.~\ref{fig:paramdom}, we display canals of parametric instability in the $(q,\gamma)$-plane for three values of the QFs parameter $\gamma$. The walls of the canals are given by Eq.~\eqref{eq:paramInstCrit}. For wavenumbers within the canals, parametric instability is expected to occur, while for wavenumbers in the neighborhood of the canals, we expect parametric stability. It can be seen that the canals, which are not straight, narrow for increasing background wavenumbers $q$, and for increasing $\gamma$.
Each of these canals actually lies in the modulationally stable part of the parameter domain, as shown in Fig.~\ref{fig:paramdom2}. Depending on the choice of parameters, the instability windows, which represents a cut at specific $\epsilon$ and $\gamma$ in Fig.~\ref{fig:paramdom}, can go from tiny in Fig.~\ref{fig:paramdom2}(a) to broader and well visible enough in Fig.~\ref{fig:paramdom2}(d).

\begin{figure}[hbt]
	\centering	
	\includegraphics[width=0.23\textwidth]{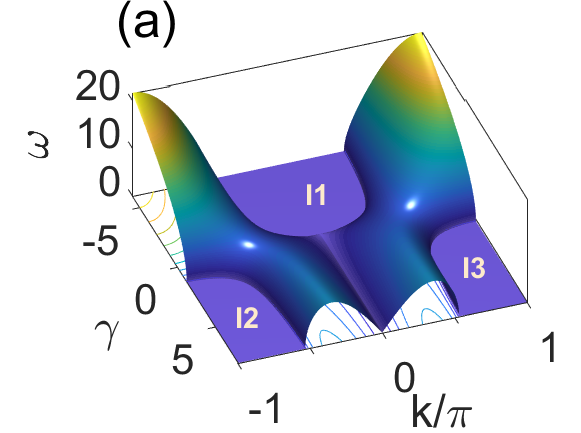}
	\includegraphics[width=0.23\textwidth]{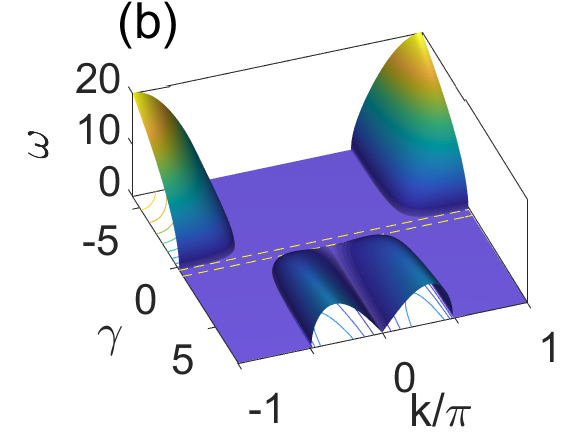}
	\caption{Separatrix sheets, which cover parametrically unstable modes in tunnels below it and keeps parametrically stable ones above it. We used (a) $g_0=1.0$ and (b) $g_0=-1.0$, with  $\eta_0=1.0$ and $u_0=1.0$. There are three stability islands I1, I2, I3 and two full instability tunnels in panel (a). The instability tunnels are constricted in panel (b), creating an isthmus of stability depicted by yellow dashed lines and four dead-end tunnels.}\label{fig:paramdom3}
\end{figure}

\subsubsection{Parametric instability suppression} 

Relation~\eqref{eq:paramInstCrit} can be helpful also for searching the suppression of parametric instability, or at least its avoidance in the system. The idea of instability suppression is to make the bandwidth as narrow as possible or even close it. The analytically predicted value of the bandwidth of parametric instability is 
\begin{equation}\label{eq:paramInstBandwidth}
	\Delta{q}_{\text{anal}} \equiv 2 q_1\epsilon=
	\frac{C_0\, \epsilon}{(C_0+\Gamma_0)\sqrt{C_m^2-C_0^2}} .
\end{equation}
In the context of periodic driving, $C_0$ cannot be cancelled. However, we can make the ratio~\eqref{eq:paramInstBandwidth} small by selecting large effective nonlinearity strengths $\Gamma_0$ and carefully sizing the system to filter permitted perturbation wave numbers. As already shown in the section above, the instability windows can be made broader or narrower.
Besides, we can completely suppress parametric instability via an appropriate choice of parameters, which we use Fig.~\ref{fig:paramdom3} to illustrate. The separatrix sheet, defined by the real part of
$\omega_c \equiv  2\sqrt{C_m(C_m+2\Gamma_0)}$ in the $(k,\gamma)$-plane, forms in Fig.~\ref{fig:paramdom3}(a) two tunnels inside which parametrically unstable modes can be picked. These modes satisfy $\omega<\omega_c$. Modes picked anywhere outside the instability tunnels, i.e. $\omega>\omega_c$ are expected to be parametrically stable. Among the latter are, for instance, high-frequency modes with repulsive DDI, which lye well above the sheet. Another way for suppressing parametric instability consists in choosing the parameters such that $\Gamma_0<-C_m/2$. In that case, we may select the DDI strength $\gamma\in ]-\frac{\eta_0-\eta_{0c}}{4|u_0|}+\frac{1}{|u_0|^2}, \, \frac{\eta_0-\eta_{0c}}{4|u_0|}[$ under the requirement $\eta_0>\eta_{0c}+2/|u_0|$, which happens to correspond to modes in Subdomain P$_2$ of Fig.~\ref{fig:InstabDigram1}. Concrete examples of these scenarios can be attained with $u_0=1$, $g_0=1$, and $\eta_0=5>2/|u_0|+\eta_{0c}=4$, i.e. repulsive contact interaction and strong quantum fluctuations, where $\gamma\in ]\frac{1}{4},\frac{3}{4}[$. A more realistic example with weaker quantum fluctuations but attractive contact interaction is gotten when $u_0=1$, $g_0=-1$, and $\eta_0=1>2/|u_0|+\eta_{0c}=0$, where $\gamma\in ]\frac{1}{4},\frac{3}{4}[$, as portrayed in  Fig.~\ref{fig:paramdom3}(b). Infinitely many examples of these kinds can be envisioned. In all these cases, there is a constriction around $\gamma=1/(2|u_0|)$ in the two instability tunnels from Fig.~\ref{fig:paramdom3}(a), which opens a stability isthmus of width $\Delta\gamma=\frac{\eta_0-\eta_{0c}}{2|u_0|}+\frac{1}{|u_0|^2}$ for any perturbation wave number $k$ as shown in Fig.~\ref{fig:paramdom3}(b). The stability isthmus both creates a $k$-independent stable zone and connects the three stability islands from Fig.~\ref{fig:paramdom3}(a).
Hence, one may suppress parametric instability at some frequencies and DDI. In contexts where we cannot completely suppress parametric instability, we can still either lessen it by shrinking the unstable bandwidth or avoid it by operating the system with parameters away from the bandwidth.

\section{Numerical experiments}\label{sec:numerics}

The analytical results obtained in the sections above are based on linear instability analysis method. Such a technique, which is based on linearization, can well predict the linear phase of the instability as well as the instability phase diagram. However, it cannot tell what happens when the nonlinear phase of the instability sets in. To validate these results and gain more insight into the dynamics of the system, we rely on direct numerical simulations, which are the focus in this section.

\subsection{Setup and space-time dynamics}

%\begin{widetext}

\begin{figure*}[hbt]
	\centering	
	\includegraphics[width=0.99\textwidth]{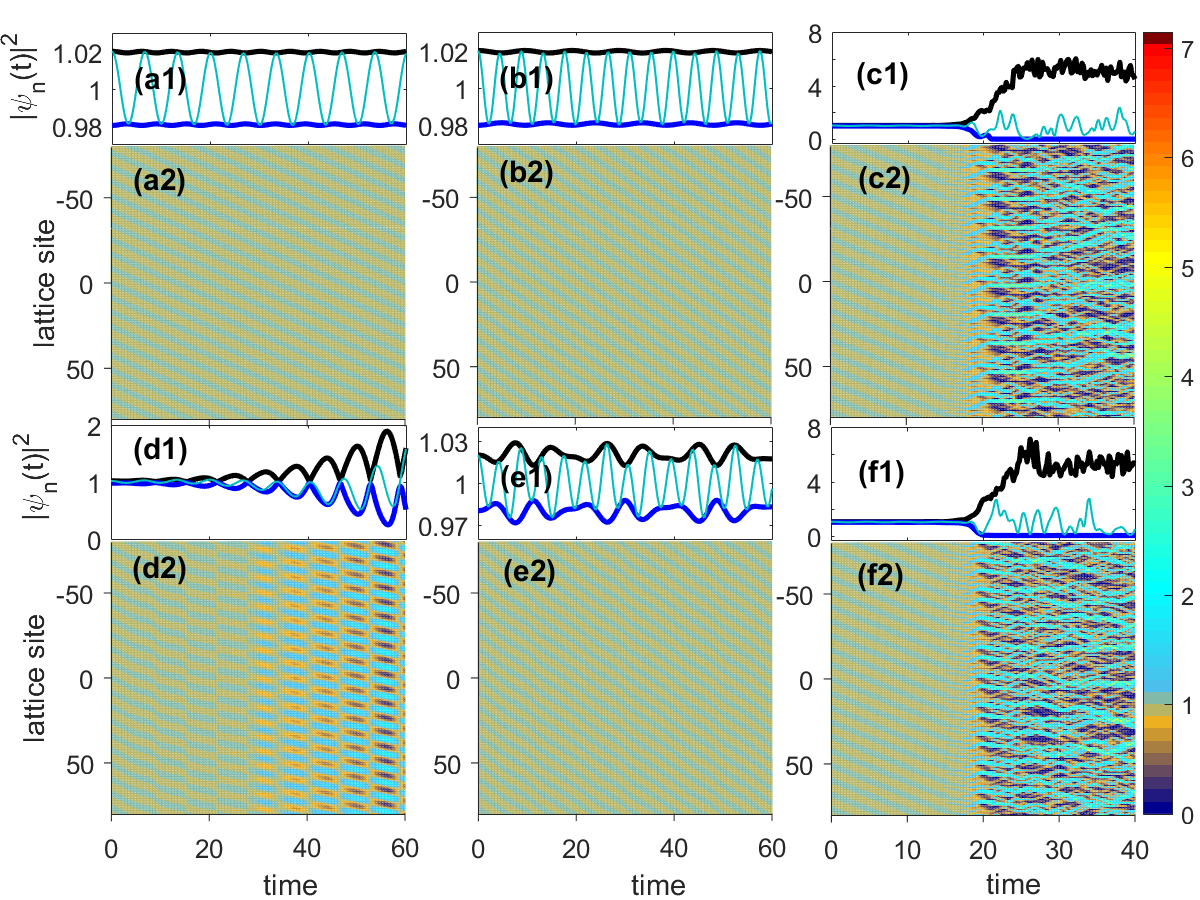}
	\caption{Spatio-temporal dynamics of the system for parametrically undriven (top two rows) and driven (bottom two rows) systems, i.e. $\epsilon=0.0$ and $\epsilon=0.10$, respectively. The first and third rows present the temporal evolution of the maximum density (black line), the minimum density (blue line), and the density at the central well (cyan line) of the lattice. The second and fourth rows are 3D contour plots of the density in space and time. We used $q=0.1006\pi$ (left column), $q=0.3008\pi$ (middle column), and $q=0.5137\pi$ (right column). The other parameters are $k=0.2354\pi$, $\gamma=-1.0$, $\eta_0=-1.0$, $u_0=1.0$, and $g_0=1.0$.} \label{fig:largeplots1}
\end{figure*}

The model being investigated numerically is the equation of motion~\eqref{eq:modelfinal}, and we use a fourth-order Runge–Kutta scheme. We choose the fixed
parameters of the system to be $\omega=1.0$ $g_0=1.0$, and a number of sites $N=2048$. The time step is $\Delta{t}=0.005$, and the initial condition is a modulated wave of the form
\begin{align}\label{eq:initcond}
	\psi_{n}(0) = u_0 \, e^{\mathrm{i} q n} (1+\varepsilon \, e^{\mathrm{i} k n} ) .
\end{align}
Let us recall in passing that the smallness parameter of perturbation $\varepsilon=0.01$ is not connected to the driving strength $\epsilon$. The only fixed parameter of the initial state is $u_0=1.0$, and we vary the wave numbers $q$ and $k$. Periodic boundary conditions are imposed to the chain of sites, leading to a quantization of the wave numbers, i.e. $q=2\pi n_1/N$ and $k=2\pi n_2/N$, where $n_1$ and $n_2$ are integers between $0$ and $N$. The solution algorithm is built upon the fourth-order Rung-Kutta scheme.

With this setup, we proceed with direct numerical simulations of Eq.~\eqref{eq:modelfinal} using the initial condition~\eqref{eq:initcond}. There are many features one can harness from the simulations. Here we focus on the evolution of the atomic density, $|\psi_n(t)|^2$, in space $n$ and time $t$.
Figure~\ref{fig:largeplots1} displays in its top two rows an illustrative spatio-temporal dynamics of the system in its static regime, i.e. $\epsilon=0.0$ for a mode picked in the subdomain M$_{\text{2b}}$ (BEC side) of Fig.~\ref{fig:InstabDigram1}. In this case, there is no periodic driving, such that an amplitude growth in the system is attributable only to modulation instability. In panels (a) and (b), for which the system is predicted to be modulationally stable (MS), the amplitude only undergo smooth and small oscillations around its average value. In panel (c), the system predicted to be modulationally unstable (MU) effectively undergo an exponential growth of its amplitude [see panel (c1)], and localized excitations are generated [see panel (c2)].
The bottom two rows of Fig.~\ref{fig:largeplots1} portray the spatio-temporal dynamics of a parametrically driven system, i.e. $\epsilon=0.10$. In this case, periodic driving leads to an amplitude growth in the system in a parametrically unstable (PU) regime, as depicted in panels (d). In a parametrically stable (PS) regime as shown in panels (e), it yields only a limited excitation, which does not change the stability of the system. These modes obviously correspond to a parameter domain where modulation instability cannot manifest itself. In the case where the system is MU, parametric driving cannot help but to keep the system nearly unstable modulationally, as shown in panels (f). That is why it is of no use to investigate parametric instability in an MU parameter domain, as we assumed in the analytical calculations. These numerical findings results well corroborate our analytical predictions.
It is worth noting that a comparison between panels (d1) and (f1) may reveal one striking feature of parametric instability. While modulation-dominated instability is characterized by a smooth or pure exponential growth of the maximum amplitude [see panel (f1)], the maximum amplitude in a parametric-dominated instability is modulated in time with  frequency $\omega$ during the exponential growth [see panel (d1)]. Indeed, these modulations have a period $T = 6.214\pm \Delta{t}$, with $\Delta{t}=0.005$ the time step, corresponding to a period $\omega \simeq 1.01$, which is nothing but the driving frequency.

\subsection{Numerical instability growth rate and parametric amplification}

In our analytical investigations, we derived expressions for the gains $\xi$ and $\xi_{\text{max}}$, as given in Eqs.~\eqref{eq:MIcond1} and \eqref{eq:Gainmax}. In this section, we are concerned by their numerical counterparts, as well as amplitude growth induced by parametric driving. 

\subsubsection{Gain estimation method}
From the linear stability ansatz and the initial condition~\eqref{eq:initcond}, the density of any MS mode is expected to oscillate between a global maximum $\text{max}_{n,t}(|\psi_n(t)|^2)= |u_0|^2(1+2\epsilon)$ and a global minimum $\text{min}_{n,t}(|\psi_n(t)|^2)=|u_0|^2(1-2\epsilon)$. In the nonlinear stage of MI, we expect the maximum squared wave amplitude $\text{max}_n(|\psi_n|^2)$ to exponentially grow well beyond $|u_0|^2(1+2\epsilon)$. A quite interesting method to get the instability growth rate consists in recording the time $\tau_s$ when a given value $\rho_s$ of the maximum amplitude over space is reached during the early exponential growth phase~\cite{Mounouna.CNSNS105_106088_2022,Wamba.PhysRevE97_052207_2018}. Then, the numerically computed instability gain can be  defined as the inverse of the time when $\text{max}_x(|\psi|^2)$ crosses the fixed value $\rho_s$, i.e.
\begin{align} \label{eq:num-gain1}
	\xi_{\text{num}} \propto \frac{1}{\tau_s}, \quad \tau_s=t\rvert_{\text{max}_n(|\psi_n|^2)=\rho_s} . 
\end{align} 
In this work, we choose to detect the wave when the perturbation strength is five times higher than initially, which corresponds to a 10\% growth of the wave density, i.e. $\rho_s=1.10 |u_0|^2$. Any other close choice would lead to similar results, at least qualitatively.

\subsubsection{Numerical gain profiles}
\begin{figure}[hbt]
	\centering	
	\includegraphics[width=0.50\textwidth]{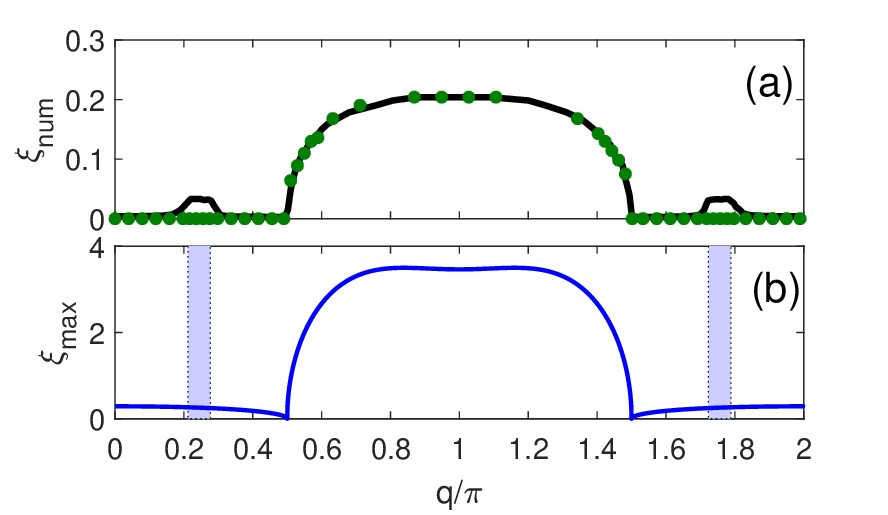}
	\caption{Numerical vs. analytical instability growth rate. Panel (a) shows the numerical instability growth rate for a parametrically driven system and $\epsilon=0.10$ (solid black line) and $\epsilon=0.0$ (green markers). We used $\omega=1.0$, $\gamma=-1.0$, $k=0.25\pi$, $u_0=1.0$, $\eta_0=-1.0$, $g_0=1.0$ and $t_{\text{max}}=70$. The bottom panel (b) provides, for comparison purpose, the corresponding analytically predicted growth rate $\xi_{\text{max}}$, which was already shown in Fig.~\ref{fig:paramdom2}(c).}\label{fig:numgain1}
\end{figure}

The numerical gain~\eqref{eq:num-gain1} allows us to successfully identify all unstable modes. An example is shown in Fig.~\ref{fig:numgain1} for modes picked in the subdomain M$_{\text{2b}}$ of Fig.~\ref{fig:InstabDigram1}. In panel (a), the green marks depict a mode of the system in absence of driving, when the zone $q=]\pi/2,3\pi/2[$ contains MU modes, while the zone $q=]0,\pi/2[ \cup ]3\pi/2, 2\pi[$ contain MS modes characterized by a zero gain. In the presence of driving, while the MU zone is kept, instability windows, characterized by two small bumps of nonzero gains, appear in the MS zones. This evidence endorses the analytical results displayed in panel (b), where grey bands represent PU windows. The agreement between the analytical predictions and the numerical results looks perfect.

\begin{figure}[hbt]
	\centering	
	%	\includegraphics[width=0.50\textwidth]{FigureTex/FigAmpliA.eps}
	%\\
	%	\includegraphics[width=0.50\textwidth]{FigureTex/FigAmpliB.eps}
	\includegraphics[width=0.50\textwidth]{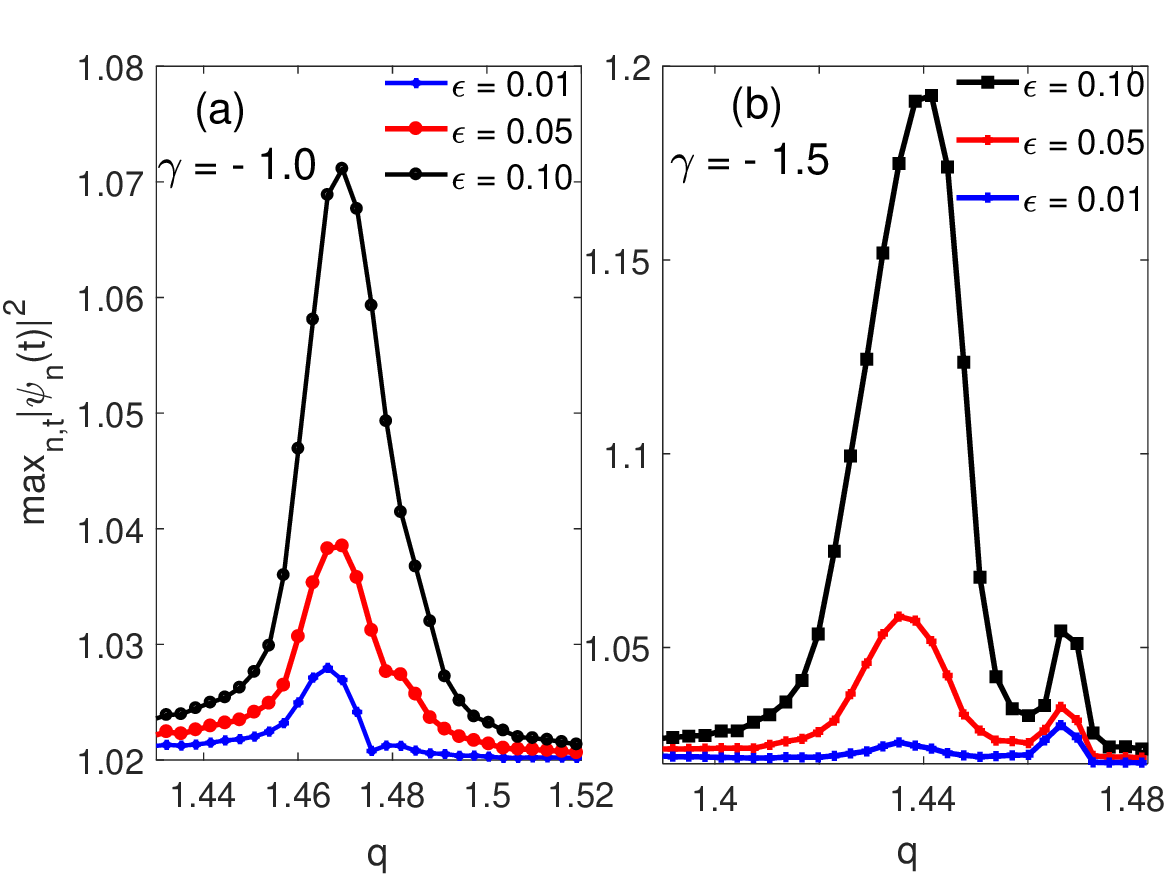}	\caption{Parametric resonances at wave numbers around $q=q_0$ for various values of $\epsilon=0.01, 0.05, 0.10$, when (a) $\gamma=-1.0$, and (b) $\gamma=-1.5$. The other parameters are $k=0.4\pi$, $\omega=1.0$, $u_0=1.0$, $\eta_0=-1.0$, $t_{\text{max}}=50$.}\label{fig:paramres}
\end{figure}

\subsubsection{Parametric amplification}
We can think of parametric amplification as a phenomenon whereby a parametric nonlinearity and a pump wave are used to amplify a signal~\cite{Trovatello.NatPhot15_6_2021}. In the context of our DNLS equation, parametric nonlinearity and the pump wave are both provided by periodic driving of the two-particle contact interaction.
Due to the driving, for PU modes and for background wavenumbers $q$ around $q_0$, the system enters a parametric resonance regime, and the wave amplitude begins to undergo a modulated exponential growth in time. To study this parametric amplification, we recorded the amplitude after a given runtime for various parameters values. As shown in Fig.~\ref{fig:paramres}(a) and (b) for two values of the DDI strength, a resonance peak is always reached around $q=q_0$ as predicted in Eq.~\eqref{eq:condq}. The peaks heighten and broaden as the driving strength $\epsilon$ is increased. In the case of quite strong DDI as portrayed in panel (b), we observe an additional small peak, which was not predicted by our analytical result. The additional peak heightens with increasing driving strengths.
From the full-width at half maximum (FWHM) of the wave amplitude around resonance, one may check the validity of our linearization. When $\gamma=-1$ like in Fig.~\ref{fig:paramres}(a), the analytical bandwidth of parametric instability~\eqref{eq:paramInstBandwidth} yields $\Delta{q}_{\text{anal}} \simeq 0.0003\pi, 0.0014\pi$ and $0.0030\pi$ for $\epsilon=0.01, 0.05$, and $0.10$, respectively.  When $\gamma=-1.5$ like in Fig.~\ref{fig:paramres}(b), it yields $\Delta{q}_{\text{anal}} \simeq 0.0005\pi, 0.0029\pi$ and $0.0057\pi$ for $\epsilon=0.01, 0.05$, and $0.10$, respectively. The numerical results based on the FWHM for $\gamma=- 1$ yield $\Delta{q}_{\text{num}} \simeq 0.0046\pi, 0.0059\pi$ and $0.0066\pi$ for $\epsilon=0.01, 0.05$, and $0.10$, respectively. We realize that the numerically computed bandwidths are broader than the ones predicted via linearization, and do not vary linearly as per our predictions. However, both the analytical and numerical results agree on the increase of the unstable bandwidth with the driving strength $\epsilon$.

\subsection{Parametric instability suppression}

\begin{figure}[hbt]
	\centering	
	\includegraphics[width=0.50\textwidth]{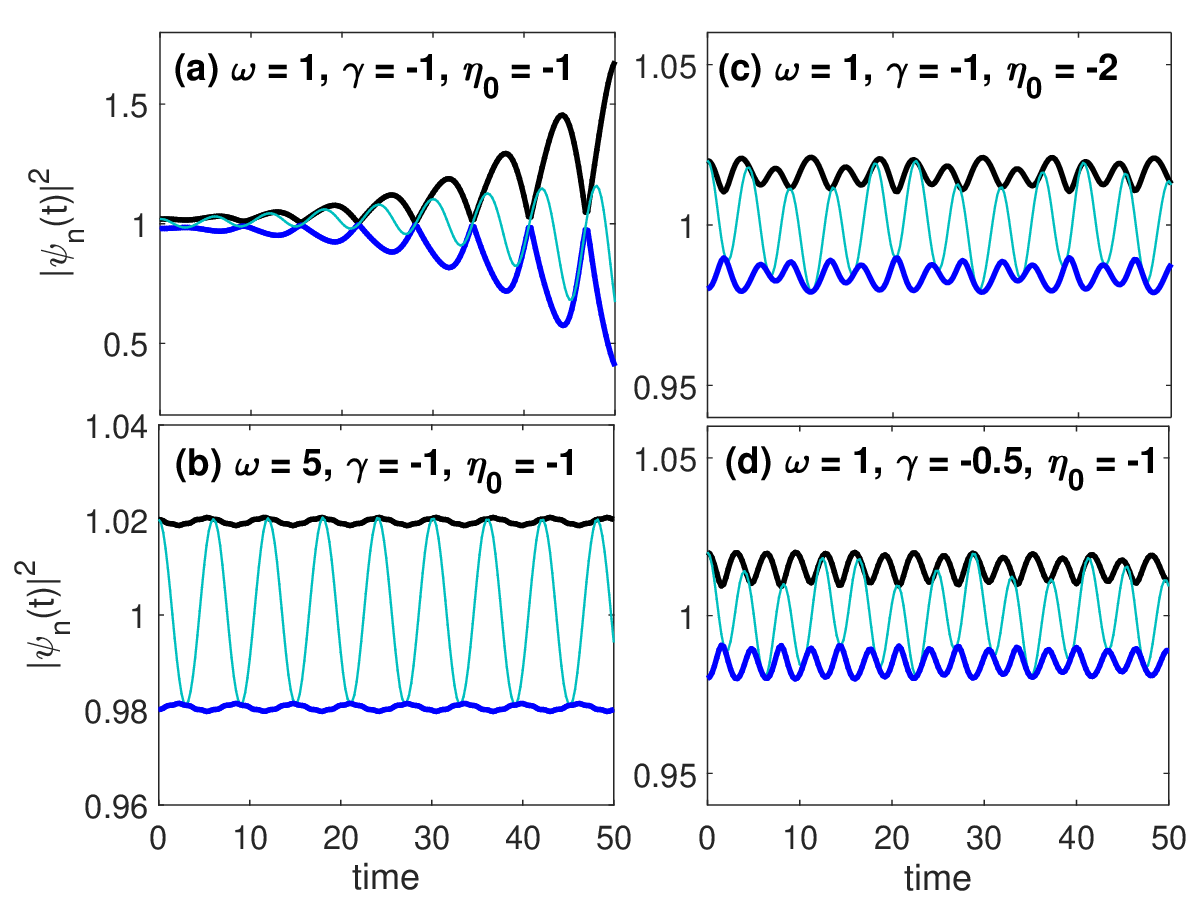}
	\caption{Suppression of parametric instability in the system. Parametric amplification (a) is suppressed using fast driving (b), stronger QFs (c), and weaker DDI (d). We used $q=0.1315\pi$,  $k=0.2354\pi$, $u_0=1.0$, and $g_0=1.0$.} \label{fig:largeplots2}
\end{figure}

In the quest to achieve suppression of parametric instability in the periodically driven system, we propose three possible scenarios. A prime method consists in periodically driving at large frequencies. As shown in Fig.~\ref{fig:largeplots2}, parametric instability, which exists at $\omega=1$, in the mode in panel (a), disappears for $\omega=5$, as portrayed in panel (b). This result, sometimes referred to as nonlinearity management~\cite{Abdullaev-Optik-228-166213-2021}, is well-known in Floquet engineering, where averaging over rapid modulations of the nonlinearity creates an effective time-independent nonlinearity.
Another method of parametric instability suppression may consist in controlling quantum fluctuations in the system. Changing the QFs strength from $\eta_0=-1$ to $-2$ as presented in Fig.~\ref{fig:largeplots2}(c) stabilizes the system against parametric amplification.
The third method for suppressing parametric instability amounts to controlling the strength of dipole-dipole interaction. For instance, changing it from $\gamma=-1$ to $-0.5$ as in Fig.~\ref{fig:largeplots2}(d) helps the system avoid parametric instability.
Our results highlight the significant impact of frequency, quantum fluctuations and long-range interactions on the stability and dynamics of the system under periodic driving. This offers insights into the control and manipulation of quantum gases in optical lattices.

\section{Conclusion}\label{sec:conclude}

In this work, we examined the dynamical instabilities of an ultracold Bose-Bose mixture with long-range dipole-dipole interactions, trapped in optical lattices and subject to periodically varying contact interaction, including effects of beyond-mean-field corrections due to quantum fluctuations. In the deep OL wells regime, we derived a discrete nonlinear Schr\"odinger equation that rules the dynamics of the system. 
Then using the linear stability analysis and multiple scale analysis, we revealed the stability diagrams in parameter spaces, including windows and canals of instability. Direct numerical calculations were conducted to endorse the analytical findings.
The roles of long-range interactions and quantum fluctuations were explored. We found that quantum fluctuations significantly attenuate the instability of the system. Our results also highlight the influence of dipole-dipole interactions on the parametric instability. We demonstrate that the anisotropy (which controls the sign) and long-range nature of these interactions can either suppress or enhance instability thresholds. In particular, repulsive dipole-dipole interaction lessens the instability. We showed that under some conditions, it is even possible to achieve a complete suppression of parametric instability.
The interplay between the contact interaction strength, the quantum fluctuations, and the dipolar interaction reveals new instability regimes that are otherwise absent in systems with purely short-range interactions.
This study not only advances our understanding of parametric instabilities in ultracold quantum gases but also provides a framework for exploring the role of long-range interactions in other quantum systems. The insights gained from this work could have implications for the design of quantum simulators and the realization of exotic quantum phases in ultracold atomic systems in optical lattices.

\newpage

%\section*{References}

\end{document}